\documentclass[a4paper]{llncs}

\usepackage{url}
\usepackage{amssymb}
\usepackage{amsmath}
\usepackage{epsfig}
\usepackage{subfigure}

\newcommand{\RR}{\bbbr}
\newcommand{\NN}{\bbbn}

\newcommand{\vs}{\vspace{0.3cm}}
\newenvironment{lists}[1]{
                 \begin{list}{}{
                     \setlength{\listparindent}{0in}
                     \settowidth{\labelwidth}{#1}
                     \setlength{\leftmargin}{\labelwidth}
                     \addtolength{\leftmargin}{\labelsep}
                     }
                 }{
                 \end{list}
                 }

\newenvironment{given-find}[2]{
                               \vs 
                               \noindent \hrule
                               \begin{lists}{Given:XX}
                               \item[\sc Given: \hfill] #1                                 
                               \item[\sc Find: \hfill] #2                               
                               \vs 
                               \noindent \hrule 
                               }{
                               \end{lists}
                               }

\newcommand{\STrue}{\mathbf{T}}
\newcommand{\SFalse}{\mathbf{F}}
\newcommand{\SUnknown}{\mathbf{U}}

\newcommand{\SErr}[1]{\mbox{err}(#1)}

\newcommand{\SAppr}[1]{\Tilde{#1}}

\newcommand{\ind}{\hspace*{0.5cm}}
\newcommand{\SFuncInv}[1]{[#1]}

\newcommand{\Cif}[1]{\mbox{\textbf{if}}\;#1\;\mbox{\textbf{then}}\;}
\newcommand{\Celse}{\mbox{\textbf{else}}\;}
\newcommand{\Creturn}{\mbox{\textbf{return}}\;}

\title{Solving Composed First-Order Constraints from Discrete-Time Robust Control}

\author{Stefan Ratschan\inst{1}\thanks{Supported by the
Austrian Science Fund FWF in the frame of the project SFB F1303} \and Luc Jaulin\inst{2}}

\institute{Research Institute for Symbolic Computation, A-4040 Linz, Austria, e-mail:~\email{stefan.ratschan@risc.uni-linz.ac.at} \and
Luc Jaulin, LISA, 62 avenue Notre Dame du Lac, 49 000 Angers, France, e-mail:~\email{jaulin@univ-angers.fr}}

\begin{document}

\maketitle

\begin{abstract}
  This paper deals with a problem from discrete-time robust control which requires the
  solution of constraints over the reals that contain both universal and existential
  quantifiers. For solving this problem we formulate it as a program in a (fictitious)
  constraint logic programming language with explicit quantifier notation. This allows us
  to clarify the special structure of the problem, and to extend an algorithm for
  computing approximate solution sets of first-order constraints over the reals to exploit this structure.  As a result
  we can deal with inputs that are clearly out of reach for current symbolic solvers.
\end{abstract}

\section{Introduction}

A discrete time system can be described by state-space equations of the form
$\vec{x}_{k+1}=f(\vec{x}_{k},\vec{u}_{k},\vec{w}_{k})$, where $k$ is the discrete time,
$f$ is a non-linear real function, $\vec{x}_{k}$ is the state vector at time $k$, $\vec{u}_{k}$ is the control vector which can be chosen
arbitrarily in a set $U_{k}$, and $\vec{w}_{k}$ is the perturbation vector which cannot be
influenced by us but remains inside a known set $W_{k}$. In this paper we deal with the
problem of computing the set of state vectors $\vec{x}_{0}$ for which we can set the
controls in such a way that future state vectors $\vec{x}_{1},\dots,\vec{x}_{n}$ will
belong to certain sets $X_{1},\dots ,X_{n}$ chosen by us. 

This problem is closely related to the problem of characterization of viability sets
involved when studying the evolution of macro-systems arising in biology, economics,
cognitive sciences, games, and similar areas, as well as in nonlinear systems of control
theory. A good reference is the book of Aubin~\cite{Aubin:91}.

When trying to solve this problem one immediately runs into constraints that contain a
large number of universally and existentially quantified variables.  In this paper we
report on ongoing research on a method for solving such constraints. This method is
already able to compute (approximate) solutions that are far too hard to compute for
current symbolic solvers (e.g., QEPCAD~\cite{Collins:91}). 

We proceed by giving a recursive formulation of the problem which can be read as a program
in a constraint logic programming language with explicit quantifier notation.  This
formulation will allow us to clarify the special structure of the resulting first-order
constraint (i.e., formula in the first-order predicate language with predicate symbols $=$
and $\leq$, function symbols $+$ and $\times$, and rational constants, all of them with their usual
interpretation).
We exploit this structure by breaking the constraint into parts, where each
part corresponds to one stage of the system, and these parts are glued together by a
syntactic entity called ``function inversion quantifier''. We show how to extend the
method of approximate quantified constraint solving
(AQCS~\cite{Ratschan:01a,Ratschan:00c}), as developed by one of the authors, to deal with
such function inversion quantifiers. This will allow us to solve non-trivial instances of
the mentioned problem from discrete-time robust control.


The resulting approach is also applicable to several other problems from discrete-time
control, and even to the general case of first-order constraints with composition
structure, that is, first-order constraints that have the form $C(f_{n}(\dots
f_{1}(\vec{x})\dots))$. Since defining objects in hierarchies is ubiquitous in the human
modeling and problem-solving process, and since such a composition structure very often is
(implicitly) encoded in programs in CLP languages, we believe that according support in
constraint solvers will become more and more important.

\section{Problem Definition}

A \emph{discrete-time system with control and perturbation} (or short a \emph{system}) is a
tuple $(f, n, X, U, W)$, where

\begin{itemize}
\item $f: \RR^{a+u+w}\rightarrow\RR^{a}$,
\item $n\in\NN$,
\item $A\subseteq\{ 1,\dots,n \}\times\RR^{a}$,
\item $U\subseteq\{ 1,\dots,n \}\times\RR^{u}$, and
\item $W\subseteq\{ 1,\dots,n \}\times\RR^{w}$.
\end{itemize}

\begin{figure}[htb]
  \begin{center}
    \input{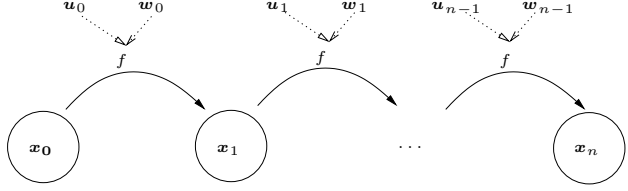}   
  \end{center}
\caption{Discrete Time System}
\label{fig:discreteSystem}
\end{figure}

The function $f$ is called \emph{transition function}, the predicate $A$ \emph{allowed
  state}, the predicate $U$ \emph{allowed control}, the predicate $W$ \emph{possible
  perturbation}.  Intuitively (see Figure~\ref{fig:discreteSystem}), such a system models
a process which starts with some value $\vec{x_{0}}$ (the \emph{initial state}) and then
applies the function $f$ $n$-times to $\vec{x_{0}}$, where at each application additional
user-definable input (control) $\vec{u}_k$ s.t.  $U(k, \vec{u}_k)$, and perturbation
$\vec{w}_k$ s.t. $W(k, \vec{w}_k)$ influence the function $f$ --- resulting in the
\emph{state at stage $k$}. In this paper we solve the \emph{problem of computing the
  robust feasible initial set}: Given a discrete-time system with control and perturbation
find an initial state such that for all following stages the state is allowed (i.e., the
  predicate $A$ holds).

Given a system $S=(f, n, A, U, W)$ we can formalize this using the following predicate,
which models the states $\vec{x}$ at stage $k$ for which we can apply input such that for
all possible perturbations the state at all future stages is allowed:
\begin{equation}
  \label{eq:program}
\begin{array}{ll}
   C_S(\vec{x}, k) \;\;:\longleftrightarrow & A(\vec{x}, k)\;\wedge\\
                                & \ind k<n \rightarrow \\
                                & \ind\ind  \exists\vec{u}\; U(\vec{u}, k)\: \wedge \\
                                & \ind\ind\ind \forall\vec{w}\; W(\vec{w}, k) \rightarrow C_{S}(f(\vec{x}, \vec{u}, \vec{w}), k+1)  
\end{array}  
\end{equation}



Our problem is to compute the robust feasible initial set $C_S(\vec{x}_0, 0)$. After expanding
this query and writing the predicates $U$ and $W$ as sets, we get:
\begin{equation}
\begin{array}{l}
A_0(\vec{x}_0)\wedge\exists \vec{u}_0\in U_0\; \forall \vec{w}_0\in W_0\\
\ind A_1(f(\vec{x}_0, \vec{u}_0, \vec{w}_0)) \wedge \exists \vec{u}_1\in U_1\; \forall\vec{w}_1\in W_1 \\
\ind\ind A_2(f(f(\vec{x}_0, \vec{u}_0, \vec{w}_0), \vec{u}_1, \vec{w}_1)) \wedge \exists \vec{u}_2\in U_2\; \forall \vec{w}_2\in W_2\\
\ind\ind\ind \dots\\
\ind\ind\ind\ind A_n(f(f(\dots f(\vec{x}_0, \vec{u}_0, \vec{w}_0), \vec{u}_1, \vec{w}_1), \vec{u}_2, \vec{w}_2))\\
\end{array}
\end{equation}

This is an expression that is definitely too hard to solve for current symbolic solvers
like QEPCAD~\cite{Collins:91} (for which the border between solvable and unsolvable
problems is around 3-5 variables), and in its raw form also for the approximate solver
developed by one of the authors~\cite{Ratschan:01a,Ratschan:00c}. Thus we have to exploit
the special structure of such a constraint in order to be able to solve it.



\section{Main Idea}
\label{sec:main-idea}

Observe that in the recursive definition of $C_{S}$ in Expression~\ref{eq:program}, we
only have a fixed number of variables in each recursion step. When expanding the
definition, each step maps the newly introduced variables into a variable vector of
lower dimension by applying the function $f$. In the other direction, when collecting
results, we can first compute the solution set of step $n$, plug the result into step
$n-1$, compute this solution set, and so on.

Here we have to solve a first-order constraint in $a+u+w$ variables for each step. But
even if we could compute the solution set of step $n$ by a symbolic solver, the size of
the result
will blow up when plugging it into the preceding steps $n-1$, $n-2$, and so on.
The reason is that the complexity of real
quantifier elimination results from the size of its quantifier-free
output~\cite{Weispfenning:88,Davenport:88}.  So we use the same idea to compute
approximate solutions instead.

For this we abstract from our concrete example and assume a constraint $C$ for which we
want to compute the solution set of $C(f(\vec{x}))$, where the dimension of $\vec{x}$ is
higher than the dimension of $f(\vec{x})$. However, we want to avoid substituting
$f(\vec{x})$ into the constraint because that would blow up the complexity by forcing us
to solve a constraint with the additional variables $\vec{x}$.


Thus we try to deal with such a situation in a special, more efficient way. For this we
write it by using \emph{function inversion quantifiers}, that is, syntactical entities of
the form $\SFuncInv{f(\vec{x})=\vec{z}}\; C(\vec{z})$.  Semantically such a constraint is
equivalent to $C(f(\vec{x}))$.  Operationally it will be handled in a different, more
efficient way. 
Now our problem reads as follows:
\begin{equation}
\begin{array}{l}
A_0(\vec{x}_{0}) \wedge \exists \vec{u}_0\in U_0\; \forall \vec{w}_0\in W_0\; \SFuncInv{f(\vec{x}_0, \vec{u}_0, \vec{w}_0)=\vec{x}_1} \\
\ind  A_1(\vec{x}_{1}) \wedge \exists \vec{u}_1\in U_1\; \forall \vec{w}_1\in W_1\; \SFuncInv{f(\vec{x}_1, \vec{u}_1, \vec{w}_1)=\vec{x}_2} \\
\ind\ind A_2(\vec{x}_{2}) \wedge \exists \vec{u}_2\in U_2\; \forall \vec{w}_2\in W_2\; \SFuncInv{f(\vec{x}_2, \vec{u}_2, \vec{w}_2)=\vec{x}_3} \\
\ind\ind\ind\ind  \dots \\
\ind\ind\ind\ind\ind A_n(\vec{x}_{n})
\end{array}
\end{equation}

Observe that in this constraint each sub-constraint beginning at a certain line has
exactly one free-variable vector $\vec{x}_{i}$. Now we can view our problem of computing
the \emph{robust feasible initial set} as the problem of solving $n$ constraints that are
glued together by function inversion quantifiers. For $1\leq i\leq n$, the $i$-th of these
constraints only contains the variable vectors $\vec{x}_{i}$, $\vec{u}_{i}$, and $\vec{w}_{i}$.

For solving these constraints we use the method for approximate quantified constraint
solving (AQCS) as developed by the first author~\cite{Ratschan:01a,Ratschan:00c}.  In the rest of the
paper we will describe how to glue together several instances of this solver by
implementing function inversion quantifiers. However, the scope of this paper will only
allow us to explain these details of AQCS that are absolutely necessary for understanding
how we deal with function inversion quantifiers.


\section{Function Inversion in Quantified Constraint Solving}
\label{sec:funct-invers-quant}

In approximate quantified constraint solving we approximate solution sets by functions
that assign ``true'' to elements that are guaranteed to be in the solution set, ``false''
to elements that are guaranteed to be out of the solution set, and that assign
``unknown'' otherwise:

\begin{definition}
  An \emph{approximate set on $B\subseteq\RR^{n}$} is a function in $B\rightarrow \{
  \STrue, \SFalse, \SUnknown \}$.
The \emph{error $\SErr{\SAppr{S}}$ of an approximate set $\SAppr{S}$ on $B$} is the volume
of the $\vec{x}$ such that $\SAppr{S}(\vec{x})=\SUnknown$. An approximate set $\SAppr{S}$
on $B$ is an \emph{approximation of a set $S\subseteq B$} iff for all $\vec{x}\in B$,
$\SAppr{S}(\vec{x})=\STrue$ implies that $\vec{x}\in S$, and $\SAppr{S}(\vec{x})=\SFalse$
implies $\vec{x}\not\in S$. 
\end{definition}
We delay the description of a concrete computer representation
of approximate sets to the end of this section.

The basic idea of the algorithm for approximate quantified constraint
solving~\cite{Ratschan:01a}, is to define for every syntactic element $p$ (e.g., $\wedge$,
$\forall$) a \emph{propagation function} $\mbox{prop}_{p}$ that computes an approximation
of the solution set of a constraint of the form $p(C_{1},\dots,C_{n})$ from approximations
of the solution set of each sub-constraint $C_{1},\dots,C_{n}$~\cite{Ratschan:00}. For
example, for the constraint $\exists x\; x^2 + y^2\leq 1$ the propagation function
$\mbox{prop}_\exists$ takes an approximation of the solution set of $x^2 + y^2\leq 1$
and returns an approximation of the solution set of $\exists x\; x^2 + y^2\leq 1$.  We
follow this approach by simply defining such a propagation function also for function
inversion quantifiers. This means that we show how to compute an approximation of the
solution set of $\SFuncInv{f(\vec{x})=\vec{z}}C(\vec{z})$ (i.e., $C(f(\vec{x}))$) from an
approximation of the solution set of $C(\vec{z})$. In a first attempt we define for an approximate
set $\SAppr{S}$ on $B$, $\mbox{prop}_{\SFuncInv{f(\vec{x})=\vec{z}}}(\SAppr{S})$ to be
$\lambda \vec{x}\!\in\!B . \SAppr{S}(f(\vec{x}))$\footnote{The notation $\lambda
  \vec{x}\!\in\!  B . f(\vec{x})$ denotes a function that takes an argument $\vec{x}\in B$
  and returns $f(\vec{x})$~\cite{Barendregt:81}.}. We will see later how to implement this
in detail.

Now a naive approach would apply the idea from Section~\ref{sec:main-idea} by computing an
approximate solution set of stage $n$, then using this information to compute an
approximate solution set of stage $n-1$, and so on (this corresponds to a computation
that follows the arrows in reverse order in Figure~\ref{fig:discreteSystem}). The problem
is that here we do not know how small we have to make the error of these approximate
solution sets in order to reach a certain desired output error. Furthermore we do not know
which parts of these approximate solution sets are needed for the end result. Thus we
compute all of them on demand.

This is also how approximate solution sets are computed within AQCS. For this it uses a function
refine that takes a constraint $C$ and a set $B$, and returns an approximation of the
solution set of $C$ on $B$ (an exact discussion of how large the error of this
approximation is allowed to be is beyond the scope of this paper). The argument $B$ is
chosen on demand within the algorithm --- it is usually just a small part of the solution
set we want to compute, and during the algorithm execution the limit of its volume goes to
zero.  So we have to extend this refinement function for the case when the outermost
symbol of $C$ is a function inversion quantifier:
\[
\begin{array}{l}
\mbox{refine}(\SFuncInv{f(\vec{x})=\vec{z}}C, B) \doteq\\
\ind \mbox{prop}_{\SFuncInv{f(\vec{x})=\vec{z}}}(\mbox{refine}(C, f(B)))\\
\end{array}
\]

The resulting overall function is recursive, the base case is reached when called with an
atomic constraint (i.e., an equality or inequality). We will deal with the problem of how
to compute $f(B)$ (i.e., the range of $f$ on $B$) later.

However, this solution does not yet suffice to solve the complexity problem. The reason is
that for each input set $B$ it computes an approximate solution set of $C$ on $f(B)$.  In
other words, it substitutes $f(\vec{x})$ into $C$ at the level of solving instead of at
the syntactic level, and we do not gain anything in terms of complexity.

This problem comes up because we never represent the approximate solution set of the
sub-constraint $C$ of $\SFuncInv{f(\vec{x})=\vec{z}}C$ explicitly.  This means that, even
if for many different $\vec{x}$ the value of $f(\vec{x})$ is the same, we recompute the
solution set of $C$ for different $\vec{x}$ that result in the same $f(\vec{x})$ by new
calls to the refinement function.  We solve this problem by remembering already computed
parts of the solution set of $C$.  We do this by wrapping the above call to the refinement
function into the following memoization function which keeps the current knowledge about
the solution set of $C$ by caching it in the global variable $\SAppr{S}$, which we
initialize by the everywhere unknown approximate solution set $\lambda
x\!\in\!\RR.\SUnknown$:
\[
\begin{array}{l}
\mbox{refineMemo}(C, B) \doteq\\
\ind  \Cif{\mbox{err}(\mbox{restr}(\SAppr{S}, B))>\varepsilon}\\
\ind\ind  \SAppr{S} \leftarrow \mbox{embed}(\SAppr{S}, \mbox{refine}(C, \mbox{choose}(\SAppr{S}, B)))\\
\ind \Creturn\mbox{restr}(\SAppr{S}, B)
\end{array}
\]
    
Here $\varepsilon$ is a pre-defined real constant, $\mbox{restr}(\SAppr{S}, B)$ returns
the restriction of the approximate set $\SAppr{S}$ (as a function) to $B$,
$\mbox{choose}(\SAppr{S}, B)$ returns a subset of $B$
for which $\SAppr{S}$ is $\SUnknown$, and $\mbox{embed}(\SAppr{S}, \SAppr{S}')$ is an
approximate set that is equal to $\SAppr{S}$ except that it is equal to $\SAppr{S'}$ on its
domain of definition.
  
Up to now we have allowed approximate sets to be arbitrary functions in $B\rightarrow \{
\STrue, \SFalse, \SUnknown \}$. For computer
representation we restrict this class to functions that are constant on finitely many
floating-point boxes. We can represent these by a set of boxes for which the approximate
set is $\STrue$, a set of boxes for which the approximate set is $\SFalse$, and a set of
boxes for which the approximate set is $\SUnknown$. Now we can easily implement the above
functions using this representation. Especially, we can implement the computation of
$f(B)$ in the refinement function by interval methods for computing an overestimation of
the range of the function $f$ on a box $B$~\cite{Neumaier:90}.

The only problem lies in the implementation of the propagation function for function
inversion quantifiers: First, the resulting approximate solution set can have an extremely
complicated structure, which can be hard or even impossible to represent by finitely many
floating point boxes.  Second, all the approximate solution sets occurring in the algorithm
have to fulfill the additional property of cylindricity~\cite{Ratschan:01a,Arnon:86} (the
clarification of this notion is beyond the scope of this paper). To solve these problems
remember that the second argument to the refinement function is a set $B$ (i.e., a box),
whose volume goes to zero during the algorithm execution. The smaller this box $B$ is, the
more probable it is that the argument to the propagation function in the refinement
function is an approximate solution set that returns the same truth value everywhere.  So
we can remove both problems easily by just returning approximate solution sets that return
the same truth value everywhere.  This means that, given an approximate set $\SAppr{S}$ on
a set $B$, we define:
\[
\begin{array}{l}
\mbox{prop}_{\SFuncInv{f(\vec{x})=\vec{z}}}(\SAppr{S}) \doteq\\
\ind\Cif{\mbox{for all}\;\vec{x}\!\in\! f(B),\;\SAppr{S}(\vec{x})=\STrue}
\Creturn\lambda\vec{x}\!\in\! B.\STrue\\
\ind\Celse\Cif{\mbox{for all}\;\vec{x}\in f(B),\;\SAppr{S}(\vec{x})=\SFalse}
\Creturn\lambda\vec{x}\!\in\! B.\SFalse\\
\ind\Celse
\Creturn\lambda\vec{x}\!\in\! B.\SUnknown
\end{array}
\]

This propagation function always computes constant approximate sets which means that it
returns the everywhere unknown function very often.  In this case AQCS will bisect the
resulting box and do further calls to the refinement function on the single pieces.  This
need for bisection instead of more intelligent box pruning methods is one of the
weaknesses of the current approach.

\section{Example}
\label{sec:examples}

We did timings on the following example:

\begin{itemize}
\item $f(x_1, x_2, u, w)= (3 x_1 x_2 + w u, 2 x_2 + x_1 )$
\item $A(x_{1}, x_{2}, t):\leftrightarrow -1\leq x_{1}\leq 1$, $-1\leq x_{2}\leq 1$
\item $U(u, t):\leftrightarrow -0.5\leq u\leq 0.5$ 
\item $W(w, t):\leftrightarrow -0.1\leq w\leq 0.1$
\end{itemize}

For $n=0$ we simply need to solve $-1\leq x_{1}\leq 1$, $-1\leq x_{2}\leq 1$. For example
for $n=3$, the total number of involved variables is $12$. We list timings for $n>0$ in
the table below. All numbers denote seconds needed on a $500$MhZ Intel Celeron PC running
Linux, where $\infty$ denotes that the run either needed more than $64$MB of memory or
more than $20$min of time.  The column with the title ``err=0.2'' shows the time for
computing an approximate solution set of error $0.2$ on the box $[ -1, 1]\times [ -1, 1]$,
the column with the title ``single'' the time for computing one single true box.  The last
column lists the time for computing an exact symbolic solution with the solver
QEPCAD~\cite{Collins:91}. For the latter we did not simply feed the input into the program
--- this would be definitely too hard to solve --- but we applied the idea of
Section~\ref{sec:main-idea} also to this case, by computing a solution of stage $n$, and
then recursively back-substituting the result for computing earlier stages. The figure
shows the computed approximate solution set for $n=2$ with error $0.2$ again on the box $[ -1, 1]\times [ -1, 1]$ (green=$\STrue$, red=$\SFalse$, white=$\SUnknown$).

\begin{figure}[htbp]
\hspace*{0.5cm}
    \begin{minipage}[t]{6cm}
    \vspace*{-4cm}
    \begin{tabular}{rrrrc} \hline
    \multicolumn{1}{c}{\hspace*{0.3cm}$n$\hspace*{0.3cm}} &
    \multicolumn{1}{c}{\hspace*{0.1cm}err=0.2\hspace*{0.1cm}} &
    \multicolumn{1}{c}{\hspace*{0.1cm}single\hspace*{0.1cm}} &
    \multicolumn{1}{c}{\hspace*{0.1cm}QEPCAD\hspace*{0.1cm}} \\ \hline
    $1$ & 0.7 & 0.0 & 18.3\\
    $2$ & 19.8 & 0.3 & $\infty$\\
    $3$ & $\infty$ & 11.9 & $\infty$\\
    $4$ & $\infty$ & $\infty$ & $\infty$\\
    \end{tabular}
    \end{minipage}
\begin{minipage}[t]{5cm}
    \epsfig{width=5cm,file=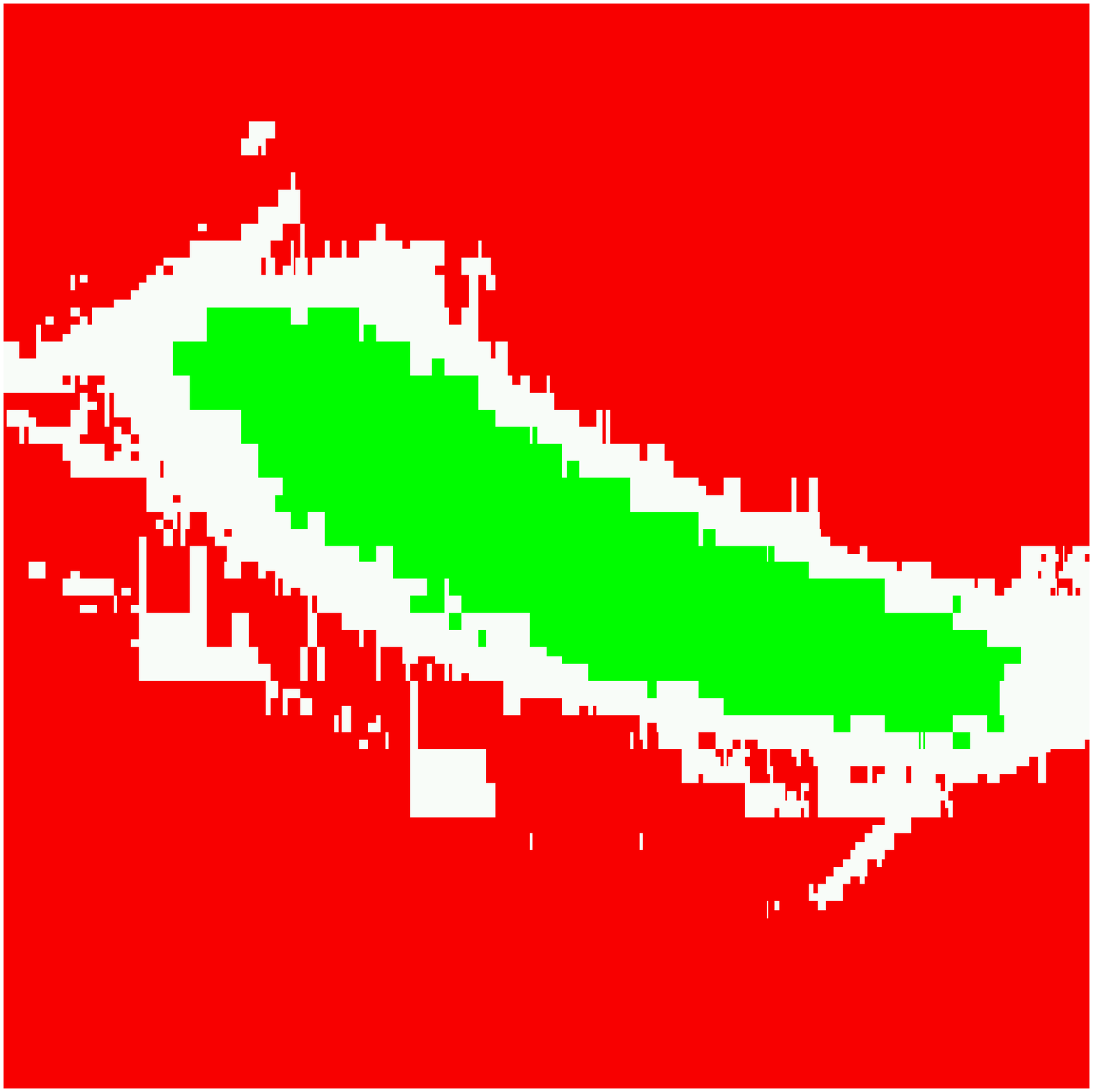}      
\end{minipage}
\end{figure}



\section{Related Work}


The behavior of various algebraic objects, such as Gr{\"o}bner
bases~\cite{Hong:96e} or subresultants~\cite{Hong:96c}, under composition is
becoming an important topic in computer algebra. The composition structure of constraints
has also been exploited for reuse of certain shared interval function evaluation in global
optimization~\cite{Kearfott:91,Hyvonen:96}, and for computing good range overestimations
of functions~\cite{Ceberio:00,Stahl:96}.

Inversion of functions on sets is done implicitly by every algorithm for solving systems
of equations~\cite{Neumaier:90} --- in this case the input set just contains one zero vector.
It is mentioned explicitly mostly for computing the solution set of systems of
inequalities~\cite{Candev:96,Walter:94}.

First-order constraints occur frequently in control, and especially robust control.  Up to
now they either have been solved by specialized methods~\cite{Ackermann:93,Zhou:97,Barmish:93}
or by applying general solvers like QEPCAD~\cite{Collins:91}. In the first case
one is usually restricted to conditions like linearity, and in the second case one suffers
from the high run-time complexity of computing exact
solutions~\cite{Weispfenning:88,Renegar:92}. We know of only one case where general
solvers for first-order constraints have been applied to discrete-time
systems~\cite{Nesic:98}, but they have been frequently applied to continuous
systems~\cite{Jirstrand:97,Dorato:95,Dorato:00}.  For non-linear
discrete-time systems without perturbations or control, interval methods have also proved to be an
important tool~\cite{Jaulin:97,Kieffer:00a}.

Apart from the method used in this paper~\cite{Ratschan:01a}, there there have been
several successful attempts at solving special cases of first-order constraints, for
example using classical interval techniques~\cite{Shary:96,Shary:99} or constraint
satisfaction~\cite{Benhamou:00}, and very often in the context of robust
control~\cite{Garloff:99,Jaulin:96,Malan:97,Vehi:00}.



\section{Conclusion}

We have designed a method for solving composed first-order constraints arising in
discrete-time robust control by extending a solver (AQCS) developed by one of the authors.
The result can solve problems that are far too hard to compute for state-of-the-art
symbolic solvers. However it is still too slow for solving problems of practical
interest.  We believe that the method is promising also in other situations where composed
first-order constraints occur.

In problems without function inversion quantifiers, AQCS spends most of the time on
computing information for atomic constraints, using tightening~\cite{Hong:94b} and range
computation. However, for inputs containing function inversion quantifiers, this is not
the case.  Usually more than 90 percent of the time is spent on propagating (usually
unknown) boxes in tasks like memo lookup or choosing boxes for further processing.
We envision two basic approaches for dealing with this situation:
\begin{itemize}
\item Try to produce less unknown boxes.
\item Try to handle the produced boxes more efficiently.
\end{itemize}

For producing less unknown boxes, the following approaches seem promising:
\begin{itemize}
\item Instead of bisection, develop a method similar to tightening~\cite{Hong:94b} or to box
  consistency methods~\cite{Hentenryck:97,Benhamou:99} on the level of function inversion
  quantifiers, for example by implementing the following specification: Given a box $B$, 
  compute a box $B'$ such that $f(B')\subseteq B$. Then the fact that $B$ is in the
  solution set of a constraint $C(\vec{z})$ implies that $B'$ is in $C(f(\vec{x}))$.
\item Pruning boxes that do not contribute to the overall
  results (in a similar way as the monotonicity test in global optimization).
\end{itemize}

For handling the produced boxes more efficiently we can:

\begin{itemize}
\item Devise strategies for choosing boxes in the context of function inversion
  quantifiers~\cite{Ratschan:01}.
\item Design efficient algorithms for the memo lookup.
\item Parallelize the method by putting each stage on a different processor.
\end{itemize}









\bibliographystyle{abbrv}

\bibliography{sratscha,sratscha_own}

\end{document}